\documentclass[floatfix,aps,prd,showpacs,amsmath,nofootinbib,preprintnumbers,twocolumn]{revtex4}

\def\half{{1\over 2}}

\def\p{\partial}

\def\half{{1\over 2}}
\def\({\left(}
\def\){\right)}
\def\[{\left[}
\def\]{\right]}

\def\e{\begin{equation}}
\def\q{\end{equation}}
\def\m{\begin{eqnarray}}
\def\n{\end{eqnarray}}

\begin{document}

\title{Field Theory at a Lifshitz Point}

\author{Bin Chen $^{1,3}$ and Qing-Guo Huang $^{2,3}$}\email{huangqg@kias.re.kr}
\affiliation{{\em $^1$ Department of Physics and State Key
Laboratory of Nuclear Physics and Technology, Peking University,
Beijing 100871, P.R.China }
\\{\em $^2$ School of Physics, Korea Institute for Advanced Study,
207-43, Cheongryangri-Dong, Dongdaemun-Gu, Seoul 130-722, Korea }
\\{\em $^3$ Kavli Institute for Theoretical Physics China,
ITP-CAS, Beijing, P.R. China}}

\date{\today}

\begin{abstract}

We construct the general renormalizable actions for the scalar field
and the gauge field at a Lifshitz point characterized by the
dynamical critical exponent $z$. The Lorentz invariance is broken
down in the UV region, but is recovered in the IR limit. Even though
the theories are UV complete,  the speed of light is related to the
momentum by $z(k/M)^{z-1}$ which can go to infinity in the UV limit
for $z\geq 2$. Since the Lorentz invariance is broken down, the
dispersion relation is modified and the time delays in Gamma-Ray
bursts can be easily explained. In addition, we also discuss the
thermal dynamics and the size of causal patch in a FRW universe for
the field theory at a Lifshitz point.

\end{abstract}

\maketitle


\section{Introduction}

 Recently Horava proposed a quantum field theory of
gravity with the dynamical critical exponent equal to $z=3$ in the
UV region \cite{Horava:2009uw}. Though the spatial isotropy is still
assumed to be kept, the isometry between space and time is got lost.
The degree of anisotropy between space and time is measured by the
dynamical critical exponent $z$, \e {\vec x}\rightarrow b {\vec
x},\quad t \rightarrow b^z t. \label{scaling} \q The theory proposed
in \cite{Horava:2009uw} describes the interaction of
non-relativistic gravitons at short distances, and recovers nearly
the Einstein's gravity in the IR region with some highly suppressed
higher-spatially-derivative modifications. Such a theory is at least
power-counting renormalizable in the $3+1$ dimensional spacetime.
Some solutions of Horava gravity theory were given in
\cite{Lu:2009em,Nastase:2009nk,Colgain:2009fe}. Since Horava gravity
has a very nice UV behavior, it has been applied to investigate the
physics in the early universe in
\cite{Takahashi:2009wc,Calcagni:2009ar,Kiritsis:2009sh,Mukohyama:2009gg,Brandenberger:2009yt,Piao:2009ax,Gao:2009bx}.
An interesting result is that the perturbation of the scalar field
with $z=3$ is scale invariant in the universe where the scale factor
goes like $a(t)\sim t^p$ with $p>1/3$ \cite{Mukohyama:2009gg}. It
may provide an alternative model to the inflation. But there are
still many open questions in this area, for example how to solve the
flatness problem without inflation. Other related works are given in
\cite{Horava:2008jf,Visser:2009fg,Horava:2008ih,Maccione:2009ju,Carvalho:2009nv,Horava:2009if,Volovich:2009yh,Kluson:2009sm,Cai:2009pe}.

In fact, the first field theory model exhibiting the above
anisotropic scale invariance (\ref{scaling}) has been known for a
long time. It is the so-called Lifshitz scalar field theory with the
critical exponent $z=2$, \cite{Lifshitz},
 \e
 {\cal L}=\int d^2x dt \((\partial_t \phi)^2 -\lambda (\Delta^2
 \phi)^2\).
 \q
It has a line of fixed points parameterized by $\lambda$. Such fixed
points with anisotropic scale invariance are usually called the
Lifshitz points.  The Lifshitz scalar field theory and its
generalizations have been used to study quantum phase transitions in
various strongly correlated electron systems \cite{Quantumphase}.
Moreover the nontrivial gauge theories with the Lifshitz fixed
points in $(2+1)$ dimension has been discussed in \cite{gaugeL}. And
in \cite{Horava:2008jf} a different construction on the non-Abelian
gauge theories with $z=2$ in arbitrary dimensions was presented.

 In this paper we temporarily forget about the gravity and
only focus on the classical field theory at a possible Lifshitz
point with arbitrary dynamical critical exponent $z$ and figure out
the most general renormalizable actions for the scalar and the gauge
fields. Due to the anisotropic scaling, the power counting of the
fields is different from the one in usual field theory. As a result,
for a field theory with $z\geq 2$, it has the marginal terms with
higher spatially derivatives and also has more renormalizable
interactions. This leads to the modification of the dispersion
relation in the UV limit. And more importantly, due to the breaking
of Lorentz invariance, the speed of light at UV may turn to
infinity. The fact that the Lorentz invariance just appears as
accidental symmetry at IR provide a natural mechanism of Lorentz
symmetry breaking. As an application, the issue of time delays in
Gamma-ray bursts could be addressed in this context.

 Our paper will be organized as follows. The general
renormalizable actions for the scalar field and the gauge field are
proposed in Sec.~2 and Sec.~3 respectively. As an application, we
provide a possible explanation for the time delays in Gamma-ray
bursts due to the modification of the dispersion relation in Sec.~4.
The thermal dynamics and the size of causal patch in a FRW universe
for the field theory at a Lifshitz point are discussed in Sec.~5 and
6 respectively. Some inspired discussions are included in Sec.~7.

\section{The renormalizable scalar field theory at a Lifshitz point in $d+1$ dimensions}

In this section we will construct the most general renormalizable
action for the scalar field theory with a dynamical critical
exponent $z$ in $d+1$ dimensions. The spacetime is assumed to be
$R\times R^d$ with the coordinates \e (t,{\vec x})\equiv (t,x^i), \q
for $i=1,2,...,d$. The spacetime metric takes the form \e
ds^2=-N^2dt^2+g_{ij}(dx^i-N^idt)(dx^j-N^jdt), \q where $g_{ij}$ are
the $d$-dimensional spatial metric of signature $(+...+)$, $N$ is
the lapse function, and $N^i$ is the shift factor.  The field theory
is assumed to have a UV fixed point with the scaling properties
given in Eq.(\ref{scaling}). In the case of general $z$, the
classical scaling dimensions of the coordinates in the unit of the
spatial momenta are \m [t]_s=-z,\quad [{\vec x}]_s=-1,\quad
[\Delta\equiv\p^i\p_i]_s=2, \n and the classical scaling dimensions
of the fields are \e [g_{ij}]_s=0,\quad [N_i]_s=z-1,\quad [N]_s=0.
\q

The prototype of a quantum field theory is the theory of a single
Lifshitz scalar $\phi(t,{\vec x})$ whose dynamics is supposed to
be governed by the following action,
 \e S=\half \int dtd^dx
N\sqrt{g}\[{1\over N^2}\(\p_t\phi-N^i\p_i\phi\)^2-\sum_{J\geq
2}{\cal O}_J\star \phi^J\], \label{rns} \q where ${\cal O}$ is an
operator which can be expanded by \e {\cal O}_J=\sum_{n=0}^{n_J}
(-1)^n{\lambda_{J,n} \over M^{2n+{d-1\over 2}J-d-1}}\Delta^n, \q
here $\lambda_{J,n}$ are the energy dimensionless coupling constant.
The $\star$ product in Eq.(\ref{rns}) contains all possible
independent combinations of $\Delta$ and $\phi$ up to a total
derivative. For example, \e \Delta^3\star \phi^3=c_1 (\Delta \phi)^3
+ c_2 (\Delta^2\phi)(\Delta \phi)\phi + c_3 (\Delta^3\phi) \phi^2,
\q where $c_1, c_2, c_3$ are the dimensionless parameters. For
simplicity we can assume $c_1=1$. Here we mainly work in the
Minkowski spacetime and then have $g_{ij}=\delta_{ij}$, $N^i=0$ and
$N=1$. From the kinetic term in the action (\ref{rns}), the scalar
field $\phi$ has the scaling dimension \e [\phi]_s= {d-z \over 2}.
\q The case of $z=d$ corresponds to a very special field theory in
which the scalar field is dimensionless and the power of $\phi$ can
be arbitrary large. The action for the scalar field with $z=d=3$ has
been written down in \cite{Kiritsis:2009sh}. In general, the scaling
dimension of the coupling constant $\lambda_{J,n}$ in the unit of
the spatial momenta is \e [\lambda_{J,n}]_s=z+d+{z-d \over 2}J-2n.
\q In order that this theory is power-counting renormalizable,
$[\lambda_{J,n}]_s$ is required to be not less than zero, namely \e
n\leq {z+d\over 2}+{z-d \over 4}J. \q Therefore \e n_J=\hbox{max}
\left\{n\in Z \left| n\leq  {z+d\over 2}+{z-d \over 4}J \right.
\right\}. \q If $z<d$, $n\geq 0$ implies $J\leq 2(z+d)/(d-z)$. If
$z\geq d$, there is no upper bound on $J$.\footnote{A similar result
was obtained in \cite{Visser:2009fg}. } For $J=2$, we have $n\leq
z$.

In the UV limit, the operator ${\cal O}_J$ is dominated by  \e (-1)^{n_J} {\lambda_{J,n_J}\over
M^{2n_J+{d-1\over 2}J-d-1}}\Delta^{n_J}, \q which takes the form
of \e \lambda_{J,n_J} {k^{2n_J}\over M^{2n_J+{d-1\over 2}J-d-1}}
\q in the momentum space, where $k=|{\vec k}|$. Therefore the
stability of the field theory in the UV limit requires that
$\lambda_{J,n_J}$ be positive.

Without loss of generality, we assume \e \lambda_{2,z}=1. \q The
effective mass term corresponds to $J=2$, namely \m
&&\half \sum_{n=0}^{z} (-1)^n{\lambda_{2,n} \over M^{2n-2}} \phi \Delta^n \phi \nonumber \\
&=&\half \sum_{2\leq n \leq z} (-1)^n{\lambda_{2,n} \over
M^{2n-2}} \phi \Delta^n \phi-\half
\lambda_{2,1}\phi\Delta\phi \nonumber \\
&+& \half \lambda_{2,0}M^2 \phi^2. \n In the IR fixed point, the
mass square is nothing but $m^2=\lambda_{2,0}M^2$ and the speed of
light is given by $c=\sqrt{\lambda_{2,1}}$. Here we assume that the
Lorentz invariance of the field theory is recovered in the IR limit,
which requires $\lambda_{2,1}=1$. Now the dispersion relation for
this field theory can be written down by \e \omega^2=m^2+{\vec
k}^2+\sum_{2\leq n \leq z}{\lambda_{2,n} \over M^{2n-2}}{\vec
k}^{2n}. \q For $z=1$, the last term in the above equation does not
exist and the standard dispersion relation is recovered. For $z\geq
2$ the dispersion relation is changed. The group velocity is given
by \e v_g={k\over \omega}\[1+\sum_{2\leq n \leq z}n\lambda_{2,n}
\left({k\over M}\)^{2n-2}\]. \q In the UV limit $(k\gg M)$, \e
v_g\simeq z \({k\over M}\)^{z-1}, \q which goes to infinity for
$k\rightarrow \infty$ if $z\geq 2$. It is not surprised because the
special relativity is broken down in the UV limit. In the IR region,
the speed of light is modified to be \e c_g= 1+{3\over
2}\lambda_{2,2}\({k\over M}\)^2 + {\cal O}\((k/M)^4\). \q If $z\geq
3$, $\lambda_{2,2}$ can be positive or negative. As long as
$\omega^2$ is positive, the field theory is always stable.

Of particular interest is the case when $z=3,d=3$. In this case,
the scalar field could couple to Horava-Lifshitz gravity and
provide an alternative to inflation. Note that in this case, the
scalar field is dimensionless and renormalizability gives no
constraint on the scalar potential $V(\phi)$.

\section{The renormalizable Yang-Mills theory at a Lifshitz point in $d+1$ dimensions}

In this section we switch to the Yang-Mills theory with an arbitrary
dynamical critical exponent $z$ in $d+1$ dimensions. The gauge field
is a one-form in $(d+1)$-dimensional spacetime, with the spatial
components $A_i=A_i^a(t,{\vec x})T_a$ and a time component
$A_0=A_0^a(t,{\vec x})T_a$. The Lie algebra generators $T_a$ of the
gauge group ${\cal G}$ satisfiy \e [T_a,T_b]=i{f_{ab}}^cT_c. \q The
Lie algebra is normalized to be Tr$(T_aT_b)=\half \delta_{ab}$. The
gauge transformations are \e \delta_\epsilon A_0=\p_t
\epsilon-i[A_0,\epsilon],\quad \delta_\epsilon
A_i=\(\p_i\epsilon^a+{f_{bc}}^a A_i^b\epsilon^c\)T_a\equiv D_i
\epsilon. \q The gauge invariant field strengths are given by \m
E_i&=&\(\p_tA_i^a-\p_iA_0^a+{f_{bc}}^aA_i^bA_0^c\)T_a \nonumber \\ &=&\p_t A_i-\p_iA_0-i[A_i,A_0],\\
F_{ij}&=&\(\p_iA_j^a-\p_jA_i^a+{f_{bc}}^aA_i^bA_j^c\)T_a \nonumber \\ &=&\p_i
A_j-\p_jA_i-i[A_i,A_j]. \n 
Since the symmetry between space and time is broken down
for $z\neq 1$, we will write the action in terms of  the electric
field strength $E_i$ and the magnetic field strength $F_{ij}$. The
engineering dimensions of the gauge field components at the
Lifshitz point are
 \e [A_0]_s=z,\quad [A_i]_s=1, \q
 and then the
engineering dimensions of the field strengths become
 \e
[E_i]_s=z+1,\quad [F_{ij}]_s=2. \q Similar to \cite{Horava:2008jf},
we choose a natural gauge-fixing condition, \e A_0=0, \quad
\hbox{and}\quad \p_iA_i=0. \q In order to keep the unitarity, the
Lagrangian should contain a kinetic term which is only quadratic in
the first time derivatives of the gauge field. Here the only choice
is Tr$(E_iE_i)$. The action in terms of  the gauge field strength
$E_i$ and $F_{ij}$ could be of the form,
 \e S=\half
\int dtd^d x
\[{1\over g_E^2}\hbox{Tr}(E_i E_i) - \sum_{J\geq 2}{\cal O}_J
\star F^J\], \q where \e {\cal O}_J={1\over g_E^J}\sum_{n=0}^{n_J}
(-1)^n {\lambda_{J,n} \over M^{2n+{d+1\over 2}J-d-1}} D^{2n}. \q
Here $F$ and $D$ are the abbreviated denotation for the field
strength $F_{ij}$ and the covariant derivative $D_k$ respectively,
and $\lambda_{J,n}$ are the coupling with zero energy dimension.
Similarly $D^{2n}\star F^J$ also contains all possible independent
combinations of $D_k$ and $F_{ij}$. Now the scaling dimensions of
$g_E$ and $\lambda_{J,n}$ are respectively given by \e [g_E]_s={z-d
\over 2}+1, \quad [\lambda_{J,n}]_s=z+d+{z-d-2\over 2}J-2n. \q The
renormalizable condition for $E_i$ is $[g_E]_s\geq 0$, namely \e
z\geq d-2. \q For $z=1$, the gauge theory is renormalizable only
when $d\leq 3$. Since there is no symmetry relating the kinetic term
and the potential terms, we still need to find out the
renormalizable conditions for the potential terms. A simple way to
work them out is to rescale the gauge field $A_i^a$ to the canonical
one ${\tilde A}_i^a$ which is related to $A_i^a$ by \e {\tilde
A}_i^a=A_i^a/g_E, \q and then the gauge field strengths become \m
{\tilde E}_i&=&\p_t {\tilde A}_i=E_i/g_E,\\
{\tilde F}_{ij}&=&\p_i {\tilde A}_j-\p_j {\tilde A}_i-ig_E[{\tilde A}_i,{\tilde A}_j]=F_{ij}/g_E.
\n
The action for the canonical gauge field is
\m
S&=& \half \int dtd^dx\[\hbox{Tr}\({\tilde E}_i {\tilde E}_i\) \right. \nonumber \\
&-&\left. \sum_{J\geq 2}\sum_{n=0}^{n_J} (-1)^n{\lambda_{J,n} \over
M^{2n+{d+1\over 2}J-d-1}}{\tilde D}^{2n}\star {\tilde F}^J\]. \n The
renormalizable conditions for the potential terms are
$[\lambda_{J,n}]_s\geq 0$ which implies \e n_J=\hbox{max}
\left\{n\in Z \left| n\leq  {z+d\over 2}+{z-d-2 \over 4}J \right.
\right\}. \q For $J=2$, $n\leq z-1$. In order to recover the $z=1$
gauge theory in the IR limit, we set $\lambda_{2,0}=1$. On the other
hand, the UV stability requires that $\lambda_{J,n_J}$ should be
positive and $\lambda_{2,z-1}$ can be set to be 1 for simplicity.

Now we can easily write down the dispersion relation for a free
gauge field theory as follows \e \omega^2=k^2\[ 1+\sum_{1\leq n\leq
z-1}\lambda_{2,n}\({k\over M}\)^{2n}\], \q where $k=|{\vec k}|$. The
group velocity is \e v_g={d\omega \over dk}={k\over
\omega}\[1+\sum_{1\leq n\leq z-1}(n+1)\lambda_{2,n}\({k\over
M}\)^{2n}\]. \q In the UV limit $(k\gg M)$, we have \e v_g\simeq
z\({k\over M}\)^{z-1}. \q The speed of light goes to infinity for
$k\rightarrow \infty$ if $z\geq 2$. In the IR regime, \e v_g\simeq
1+{3\over 2}\lambda_{2,1} \({k\over M}\)^2. \label{spdl} \q Here a
negative $\lambda_{2,1}$ is allowed as long as $\omega^2$ is
positive definitely for $z\geq 3$. In the next section the above
modified speed of light can be used to explain the time delays in
Gamma-ray bursts.

\section{An explanation for the time delays in Gamma-Ray bursts}

Recently the Fermi LAT and Fermi GBM collaborations reported that
the photon with energy $E_h=13.22_{-1.54}^{+1.70}$ GeV arrived at
the Earth is $16.54$ s later than the low-energy photon from GRB
08916C with measured redshift of $z'=4.35\pm 0.15$
\cite{grb}\footnote{In this paper, we use $z'$ to denote the
redshift.}. If the high-energy photon was emitted at the same time
as the low-energy photon, this delay may encode the information of
Lorentz symmetry violation
\cite{Ellis:2002in,Ellis:2005wr,Jacob:2008bw}. In
\cite{Ellis:2002in,Ellis:2005wr,Jacob:2008bw}, the dispersion
relation of the photon is proposed to be modified by the effect of
quantum gravity. Some other possible explanations were suggested in
\cite{Li:2009tt,Li:2009mt}. In Sec.~4 we saw that the dispersion
relation and the speed of light of the photon field at a Lifshitz
point was modified. This fact suggests a natural way to explain the
time delays in Gamma-ray bursts.

Here we would like to give a general discussion about the time
delays in the Gamma-ray bursts. Assume that the velocity of the
photon with physical momentum $k$ is given by \e c_g(k)=1+\lambda
\(k\over M\)^\alpha. \label{cgl} \q This deformed velocity of the
photon implies that the the simultaneously emitted photons from the
source of the Gamma-ray bursts reach the Earth at different times.
In the FRW universe, the momentum of the photon is redshifted by the
expansion of the universe. The scale factor is related to the
redshift factor $z'$ by $a=(1+z')^{-1}$ and the speed of light at
the time of $z'$ becomes \e c_g(k,z')=1+\lambda \(k/a\over
M\)^\alpha=1+\lambda (1+z')^\alpha \({k\over M}\)^\alpha. \q The
comoving distance between the source of the Gamma-ray burst and the
Earth is $x_c$ which is given by \e
x_c=\int_{t_\gamma}^{t_k}c_g(k){dt\over a}, \q where $t_\gamma$ is
the time when the photon was emitted. If the high-energy and
low-energy photons were emitted at the same time $t_\gamma$, the
time delay can be easily obtained, \e \delta t \equiv
t_{k_h}-t_{k_l}\simeq -\lambda {\delta k^\alpha \over
M^\alpha}\int_0^{z_\gamma} {(1+z')^\alpha \over H(z')} dz', \q where
\e \delta k^\alpha\equiv k_h^\alpha-k_l^\alpha. \q For $\Lambda$CDM
model, we have \e
H(z')=H_0\sqrt{\Omega_m^0(1+z')^3+\Omega_\Lambda^0}, \q where $H_0$
is the present Hubble parameter, and then \e \delta t \simeq
-\lambda H_0^{-1} {\delta E^\alpha \over M^\alpha}\int_0^{z_\gamma}
{(1+z')^\alpha \over \sqrt{\Omega_m^0(1+z')^3+\Omega_\Lambda^0}}
dz', \q here $E\simeq k$ is the photon energy measured on the Earth
and $\delta E^\alpha\simeq E_h^\alpha$. In order to explain the time
delays, $\lambda$ should be negative. Here $H_0^{-1}$ is roughly the
same as the age of the universe, but $\delta t$ is only about
$16.54$ s, and hence $M$ should be much larger than $E_h$ if
$|\lambda|$ is not so small. The WMAP 5yr data \cite{Komatsu:2008hk}
indicates that $H_0=70.5$ km s$^{-1}$ Mpc$^{-1}$,
$\Omega_\Lambda^0=0.726$ and $\Omega_m^0=0.274$. Taking
Eq.(\ref{spdl}) into account, we have $\lambda={3\over
2}\lambda_{2,1}$ and $\alpha=2$. For $E_h=13.22$ GeV and $\delta
t=16.54$ s, we get \e M \simeq 60 |\lambda_{2,1}|^\half {1 \over
\sqrt{H_0 \delta t}} \ \hbox{GeV} \simeq 9.8\times 10^9
|\lambda_{2,1}|^\half \ \hbox{GeV}. \q Usually it is expected that
$|\lambda_{2,1}|\sim {\cal O}(1)$ and then a conservative estimation
of $M$ is roughly not lower than $10^{10}$ GeV. It would be better
to take this result as the constraint on the scale of Lorentz
symmetry breaking in the Lifshitz gauge field theory, taking into
account of the fact that there exist possible astrophysical sources
accounting for the time delays of the Gamma-ray bursts.

Before closing this section, we need to stress that a negative $\lambda$ in Eq.(\ref{cgl}) implies
that the theory becomes unstable and ill-defined in the UV region.
However, for a field theory at Lifshitz point with $z\geq 3$ it is
a UV  well-defined field theory which can easily explain the time
delays in Gamma-ray bursts.

\section{The thermal dynamics of the field theory at the Lifshitz point}

It would be interesting to study the thermal dynamics of the above
field theories at Lifshitz point. From the discussions in Sec.~2 and
3, the dispersion relations for both the scalar field and the gauge
field are given by \e \omega^2=m^2+k^2+...+{k^{2z}\over M^{2z-2}}.
\q The energy density at finite temperature $T$ is \e \rho\sim
\int_0^\infty \omega e^{-\omega/T} k^{d-1}dk. \q In the high
temperature limit $(T\gg M)$, the dispersion relation can be
simplified to be $w\simeq k^z/M^{z-1}$, and hence \e \rho\sim
M^{(z-1)d/z}T^{1+d/z}. \q Similarly,  the entropy density is found
to be \e s\sim M^{(z-1)d/z}T^{d/z}. \q The above scaling behaviors
imply that the field theory seems living in a $d_s=1+d/z$
dimensional spacetime. For $z=d$, $\rho\sim M^{d-1}T^2$ and $s\sim
M^{d-1}T$. We see that the thermal behaviors of these field theories
are quite different from the ones of a relativistic field theory.

We are also interested in the equation of state of matter at the
Lifshitz point with $z$ in a FRW universe. Considering $[E]_s=z$ and
the spatial volume has dimension $[V]_s=-d$, we have \e
[\rho]_s=[{E\over V}]_s=z+d. \label{rhoz} \q The metric of a FRW
universe is \e ds^2=-dt^2+a^2(t) d{\vec x}^2. \q Taking Eq.
(\ref{rhoz}) into account, we have \e \rho\propto a^{-(z+d)}. \q In
the FRW universe, the energy density of matter with the equation of
state $w$ goes like $\rho\sim a^{-d(1+w)}$. Therefore the equation
of state of matter at the Lifshitz point with $z$ is \e w={z\over
d}. \q Obviously, when $z=1$, this is exactly the equation of state
of relativistic matter in a FRW universe. Since the energy density
$\rho\propto T^{1+d/z}$, $T\propto a^{-z}$. The temperature of the
radiation with $z>1$ decreases faster than that of the relativistic
matter in an expanding universe. On the other hand, the entropy
density $s\propto T^{d/z}$ and then $s\propto a^{-d}$. This is
reasonable because the entropy density is inversely proportional to
the physical volume $a^d$.

\section{The size of causal patch for the field theory at the Lifshitz point in the FRW universe}

In this section we will figure out a new length scale $L_H$ which
characterizes the proper size of a causal patch in space for the
perturbation mode with physical momentum $p$. Consider two
particles separated by a distance $L_c$ in the comoving
coordinates at the time $t$ in a flat FRW universe. The proper
distance between them is nothing but \e L=a(t)L_c. \q If the
spatial comoving coordinates of these two particles remain
unchanged, the relative speed between them due to the expansion of
the universe is \e {dL\over dt}=\dot a L_c=HL. \q On the other
hand, the propagation velocity of the message between these two
particles through the field with dynamical critical exponent $z$
is $c_g\sim p^{z-1}/M^{z-1}$. Therefore the size of the causal
patch $L_H$ satisfies \e HL_H\sim p^{z-1}/M^{z-1}. \q At the time
when the perturbation mode stretches outside its causal patch, we
have $p\sim 1/L_H$ and then we obtain \e L_H\sim
(M^{z-1}H)^{-1/z}. \q For $z=1$, $L_H$ is nothing but the Hubble
length. For $z=d=3$, it is the same as the one found in
\cite{Mukohyama:2009gg}.

For a $(d+1)$-dimensional FRW universe dominated by the matter with the equation of state $w$, the scale factor grows up as
\e
a(t)\sim t^{2\over d(1+w)}.
\q
If $w<-1+2/d$, the universe is in an inflationary phase.
The Hubble parameter decreases as $1/t$ if $w>-1$. In order that a perturbation mode is generated within the causal patch and stretches outside the horizon in the future, we should have $a(t)>L_H(t)$ for a sufficient large t, which implies
\e
w<w_c=-1+{2z\over d}.
\q
For $z=1$, a causally generated quantum perturbation can stretch outside its causal patch and be frozen to be a classical perturbation only in an inflationary universe. But for $z>1$, it can happen even in a non-inflationary universe. Since the scalar field with the dynamical critical exponent $z$ has dimension ${d-z\over 2}$, the perturbation of such a scalar mode with $z=d$ is expected to be scale-invariant even in a non-inflationary universe. That is why ones claim that the inflation is not necessarily required when the field theory at a Lifshitz point with $z=3$ is called for in our $(3+1)$ dimensional universe. However, even though the horizon problem in hot big bang model might be released due to the super-luminosity in the UV region,  the flatness problem can be solved only in an inflationary universe. It is premature to claim that the Lifshitz field/gravity provides an alternative model to inflation.

\section{Discussions}

In this paper we constructed the most general power-counting
renormalizable actions for the scalar field and the gauge field
without considering the detailed balance condition. These field
theories at long distance reduce to the field theories with the
Lorentz invariance intact, but the symmetry between space and time
is broken down at short distance for $z\geq 2$.  Since only the
kinetic terms which is quadratic in the first time derivatives are
included, the field theories are still unitary. In this paper we
assumed that the space is isotropic. One can generalize them to the
cases with anisotropic space. Here we only proposed that the spatial
derivative operators like $\Delta^n$ appearing in the action, where
$n$ is an integer. Maybe some terms with fractional power of the
differential operator $\Delta$ could be included as well
\cite{Calcagni:2009ar}. But the physical meaning of these terms is
not well-understood.

In the original Lifshitz scalar field theory and its generalization
to non-Abelian gauge field and
gravity\cite{Horava:2008ih,Horava:2008jf,Horava:2009uw}, one may
impose the detailed balance condition to fix the potential. In these
cases, the ground state wavefunction of the theory reproduce the
partition function of a relativistic theory in lower-dimension. This
fact may suggest that the theories with the detailed balance
condition have quantum critical points. In this paper, for
generality, we did not impose any kind of the detailed balance
condition. As a result, even if we only consider the interaction
terms with the marginal dimension, the theory is just classically
scale invariant and may not be scale invariant quantum mechanically.
Obviously a careful investigation of RG flow and quantum criticality
would be a very interesting issue.

In \cite{Kachru}, the gravity duals of the anisotropic scale
invariant field theory have been constructed. It would be
interesting to investigate the gravity duals of the theories
presented in this paper.

\vspace{1.4cm}

\noindent {\bf Acknowledgments}

The work was partially supported by NSFC Grant No.10535060,
10775002 and NKBRPC (No. 2006CB805905).



\end{document}